\documentclass[aps,prl,reprint,groupedaddress,amsmath]{revtex4-1}

\usepackage{graphicx}
\usepackage{bm}
\usepackage{hyperref}

\usepackage{subcaption}

\usepackage{amsmath}
\usepackage{amssymb}
\usepackage{cancel}
\usepackage{subcaption}
\usepackage{comment}
\usepackage{slashed}
\usepackage{physics}
\usepackage{tikz-cd}

\begin{document}

\title{Onset of Quantum Chaos in Random Field Theories}

\author{Micha Berkooz}
\email[]{micha.berkooz@weizmann.ac.il}
\author{Adar Sharon}
\email[]{adar.sharon@weizmann.ac.il}
\author{Navot Silberstein}
\email[]{navot.silberstein@weizmann.ac.il}
\author{Erez Y. Urbach}
\email[]{erez.urbach@weizmann.ac.il}

\affiliation{Department of Particle Physics and Astrophysics, \\ Weizmann Institute of Science, Rehovot 7610001, Israel}

\date{\today}

\begin{abstract}
We study the quantum Lyapunov exponent $\lambda_L$ in theories with spacetime-independent disorder. We first derive self-consistency equations for the two- and four-point functions for products of $N$ models coupled by disorder at large $N$, generalizing the equations appearing in SYK-like models. We then study families of theories in which the disorder coupling is an exactly marginal deformation, allowing us to follow $\lambda_L$ from weak to strong coupling. We find interesting behaviors, including a discontinuous transition into chaos, mimicking classical KAM theory.
\end{abstract}


\maketitle

\section{Introduction}

Systems with disorder are ubiquitous in nature, and display a wide range of interesting physical phenomena. Disorder can sometimes be modeled by introducing random couplings, and averaging over these random couplings can lead to simplifications which allow for exact computations, at least when the number of degrees of freedom is parametrically large.
A notable example is the Sachdev–Ye–Kitaev (SYK) model, in which (nearly) conformal symmetry is restored at low energies  \cite{PhysRevLett.70.3339,KitaevTalk,Maldacena:2016hyu,Kitaev:2017awl}, and such averages allow for the computation of the quantum chaos exponent $\lambda_L$. The latter is defined by the fastest growing exponential mode in the double-commutator $\langle [U(t),W(0)]^2\rangle_\beta\sim e^{\lambda_L t}$ for generic operators $U,W$, over an appropriate time scale, with temperature $1/\beta$ \cite{Larkin1969QuasiclassicalMI}.

The SYK model consists of $N$ free quantum mechanical fermions deformed by a relevant all-to-all spacetime-independent disordered interaction. Generalizations have appeared which consist of $N$ copies of other free theories with similar interactions
\cite{Murugan:2017eto,Fu:2016vas,Peng:2018zap,Bulycheva:2018qcp,Chang:2021fmd,Popov:2019nja,Gross:2017vhb,Lian:2019axs}. In this letter we study quantum chaos in a more general setting by studying $N$ copies of a general core model $\mathcal{Q}$ (which can be a quantum field theory or a spin system), deformed by a spacetime-independent all-to-all interaction. In particular, we eventually focus on $Q$ being a conformal field theory (CFT) in 0+1 or in higher dimensions, and with an exactly marginal disorder interaction. In such a setup we have better control over the space of couplings over which we are averaging, eliminating the complications of the renormalization group (RG) and without having to resort to strong coupling.

The theories we study are of the form
\begin{equation}\label{eq:disordered_CFTs}
	\mathcal{Q}^N+\sum_{i_1\neq ...\neq i_q}^N J_{i_1...i_q}\mathcal{O}_{i_1}...\mathcal{O}_{i_q}\;,
\end{equation}
where $\mathcal{Q}^N$ denotes $N$ decoupled copies of the model $\mathcal{Q}$, $\mathcal{O}_i$ with $i=1,...,N$ are the $N$ copies of a local operator $\mathcal{O}$ in $\mathcal{Q}$, and $J_{i_1...i_q}$ are Gaussian random variables with variance \begin{equation}
    \langle J_{i_1...i_q}^2\rangle= \frac{J^2(q-1)!}{N^{q-1}}\;.
\end{equation} 
In the case where $\mathcal{Q}$ is a CFT, we will take $\mathcal{O}$ to be a primary operator of this core CFT, and equation \eqref{eq:disordered_CFTs} should be interpreted in conformal perturbation theory in $J$. We call such theories disordered CFTs; the simplest example of this setup is the SYK model itself, as a disordered free fermion theory. 

We will first derive self-consistency equations for the two- and four-point functions of $\mathcal{O}_i$ at leading order in $1/N$ for general disordered CFTs. These extend known results for disordered free theories like the SYK model \cite{Maldacena:2016hyu,Kitaev:2017awl}), and for spin systems \cite{BrayMoore}. We will also discuss similar results for the double-commutator (defined in \eqref{eq:double_comm} below). Although these self-consistency equations are complicated, they are tractable in perturbation theory in $J$ and allow us to establish the existence of a kernel structure from which we can extract the chaos exponent $\lambda_L$, which is the rate of growth of a double commutator. The latter is given by the fastest growing eigenvector (with eigenvalue 1) of a specific integral kernel $K_R$ (see equation \eqref{eq:double_commutator_kernel}), as long as this rate of growth is indeed positive \cite{KitaevTalk2,KitaevTalk,Larkin1969QuasiclassicalMI}.  We will denote by $\lambda_L^{\text{ker}}$ the putative chaos exponent read from the diagonalization $K_R$.

\begin{figure}
	\centering
	\includegraphics[width=0.8\linewidth]{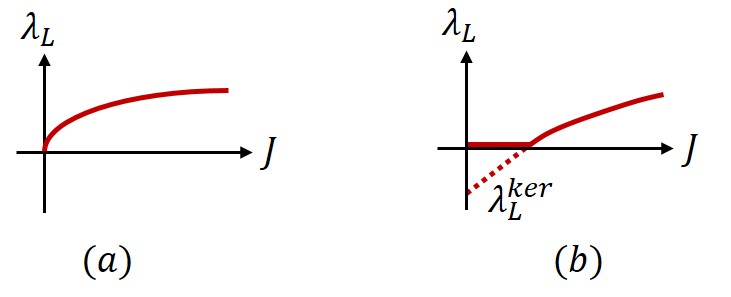}
	\caption{The behaviors we find for the chaos exponent as a function of an exactly marginal disorder deformation $J$: (a) continuous and (b) discontinuous. $\lambda_L^{\text{ker}}$ corresponds to the dashed line and $\lambda_L$ to the solid line.}
	\label{fig:discontinuouschaos}
\end{figure}
 
We will be interested in computing the chaos exponent $\lambda_L$ as a function of $J$, as it is varied from weak to strong coupling. The diagonalization of $K_R$ is a difficult process,
which can usually be done only when conformal invariance is restored. 
Normally, the disorder is a relevant deformation. To compute the chaos one first flow to the IR CFT (which is equivalent to take $J=\infty$) and find the chaos exponent there \cite{Murugan:2017eto}. 
In this work we focus on cases where the averaged interacting theory is conformally invariant for all $J$.
This can be done by demanding that the disorder interaction itself be exactly marginal, in which case the theory is conformally invariant for every realization of the couplings, and the space of $J$'s forms a conformal manifold \footnote{It is important that $\mathcal{O}$ itself does not obtain an anomalous dimension as a function of $J$, which will be the case in the examples in this letter.}. The disorder average is then simply a non-uniform average over this conformal manifold, with no RG-related complications (for recent discussions of averages over conformal manifolds see \cite{Maloney:2020nni,Afkhami-Jeddi:2020ezh}, and for the marginal case see \cite{Berkooz:2017efq}).

Surprisingly, we will find models where $\lambda_L^{\text{ker}}(J)$ is negative at weak coupling. This signals a breakdown of the approximations involved in the computation, and so in this case 
the true chaos exponent simply vanishes, $\lambda_L= 0$.
In other words, we have
\begin{equation}
    \lambda_L=\max(0,\lambda_L^{\text{ker}})\;.
\end{equation}
As a result, there are two possible behaviors for the onset of chaos: either the theory undergoes a continuous transition into chaos as in figure \hyperref[fig:discontinuouschaos]{1a} or a discontinuous transition as in figure \hyperref[fig:discontinuouschaos]{1b}, corresponding to whether $\lambda_L^{\text{ker}}$ is non-negative or negative at small enough $J$ respectively.\footnote{We implicitly assumed that the chaos exponent increases monotonically with the strength of the disorder $J$. As we start with N decoupled systems it is reasonable, but not necessarily true. For example, an integrable deformation keeps the exponent at zero, but these are hard to come by in this context.}

The discontinuous transition into chaos is a surprising result, and it is tempting to compare it to similar results in classical chaos, the most famous one being the KAM theorem. In order to sharpen the comparison, we also discuss what a single core CFT should obey in order for the transition into chaos to be discontinuous. Similar works on the onset of quantum chaos include \cite{Stanford:2015owe,deMelloKoch:2019ywq,Chowdhury:2017jzb,Steinberg:2019uqb,Maldacena:2016hyu,Peng:2018zap}, and \cite{Lian:2019axs,Hu:2021hsj} are especially relevant. 

We will apply our formalism to two classes of examples where the disorder interaction is exactly marginal. The first class is disordered generalized free fields in $1$d (following \cite{Gross:2017vhb}) and in $2$d. The second class is the disordered $\mathcal{N}=2$ supersymmetric (SUSY) $A_{q-1}$ minimal models. In practice, we will only discuss the simplest case of the $A_2$ minimal model here. We will find a discontinuous transition in the former and a continuous transition in the latter. 

More details and discussions on the computations and results can be found in the companion paper \cite{Berkooz:2021ehv}.

\section{Disorder around a nontrivial CFT}

\subsection{The kernel structure of the four-point function}

\begin{figure}
	\centering
	\includegraphics[width=1\linewidth]{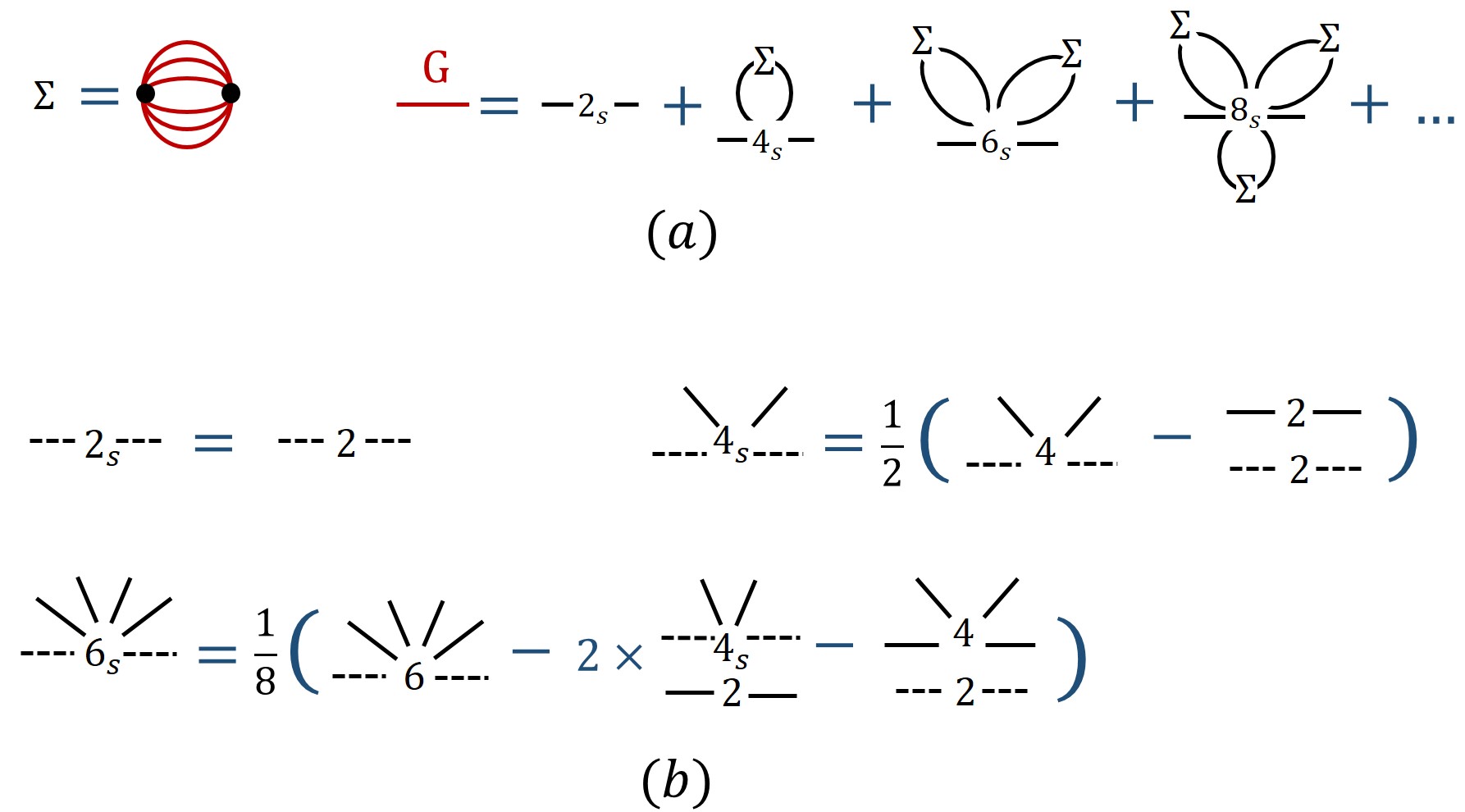}
	\caption{(a) The SD equations. $G$ insertions appear as red lines, and black dots denote insertions of the deformation \eqref{eq:disordered_CFTs}. (b) Examples of $n_s$, The subtracted $n$-point functions. Dashed lines connect to the external points, and numerical factors indicate symmetry factors. The blue numbers inside the brackets denote possible ways of permuting the legs. 
	\label{fig:sdeqsfull}
	}
\end{figure}

We start by writing a self-consistency equation for the averaged two-point function of \eqref{eq:disordered_CFTs},
\begin{equation}
	G(x)= \langle \mathcal{O}_i(x)\mathcal{O}_i(0) \rangle\;.
\end{equation}
Using the $G-\Sigma$ formalism \cite{Maldacena:2016hyu,KitaevTalk}, it can be shown that $G$ obeys a generalized Schwinger-Dyson (SD) equation at leading order in $1/N$,  which appears diagrammatically in figure \hyperref[fig:sdeqsfull]{2a}.
The equation includes subtracted $n$-point functions denoted by ``$n_s$'', which are combinations of the standard core CFT $n$-point functions with additional theory-independent subtractions which can be derived order-by order in $n$ \cite{Berkooz:2021ehv}. The first few subtracted $n$-point functions (assuming $\mathcal{O}_i$ are real) are shown in figure \hyperref[fig:sdeqsfull]{2b}. Generalizations to complex $\mathcal{O}_i$ exist, and also to superfields (in which case the diagrams correspond to supergraphs).

\begin{figure}
	\centering
	\includegraphics[width=0.9\linewidth]{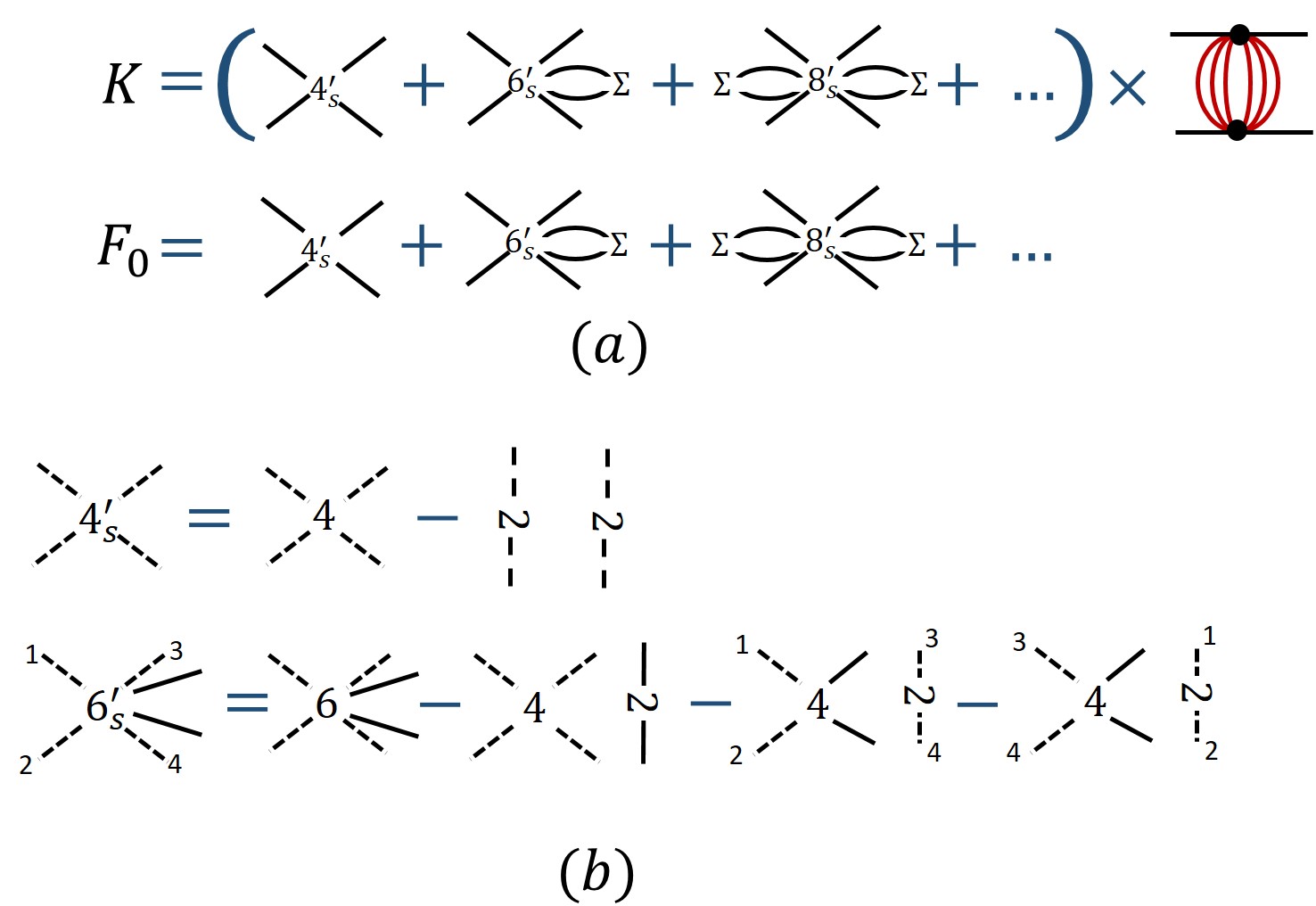}
	\caption{(a) The kernel $K$ and initial diagram $F_0$ for general disordered CFTs. Red lines denote full propagators $G$, and black dots denote insertions of the disorder interaction, with $q-2$ red propagators between each pair. (b) Examples of correlation functions $n_s'$. Dashed lines corresponds to external points, while solid lines are connected via $\Sigma$'s defined in figure \hyperref[fig:sdeqsfull]{2a}.}
	\label{fig:fullkernel}
\end{figure}

The contributions to the averaged connected four-point function \begin{equation}\label{eq:conn_four}
C=\frac{1}{N^2} \sum_{i,j}(\langle\mathcal{O}_i\mathcal{O}_i\mathcal{O}_j\mathcal{O}_j\rangle-\langle \mathcal{O}_i\mathcal{O}_i \rangle\langle \mathcal{O}_j\mathcal{O}_j \rangle)
\end{equation}
also have a simple form, and obey an iterative ladder structure similar to the case of disordered free fields:
\begin{equation}
C=\sum_{n=0}^\infty K^n F_0\;,
\end{equation}
where the kernel $K$ and the initial diagram $F_0$ are defined in figure \hyperref[fig:fullkernel]{3a}. The definition requires new subtracted $n$-point functions called $n_s'$, which are again theory-independent \cite{Berkooz:2021ehv}. The first few $n_s'$ for real $\mathcal{O}_i$ appear in figure \hyperref[fig:fullkernel]{3b}.

Some comments are in order. First, although the kernel $K$ is very complicated, knowing that an iterative ladder structure exists for the four-point function is already an important result since it allows for a systematic computation of $\lambda_L$, as we now discuss. 
Second, although a full solution of the two- and four-point functions requires knowing all $n$-point functions of the core CFT, a solution to order $J^{2n}$ in perturbation theory in $J$ only requires knowing the  $2m$-point functions for $m\leq n+1$.

\subsection{The double-commutator and chaos}

A similar analysis also applies to the computation of the double-commutator:
\begin{equation}\label{eq:double_comm}
\begin{split}
    W_R(t_1,t_2) = \frac{1}{N^2} \sum_{i,j=1}^N &\langle [\mathcal{O}_i(\beta/2) ,\mathcal{O}_j(\beta/2+i t_2)] \\
    &\cdot [\mathcal{O}_i(0), \mathcal{O}_j(i t_1)]
    \rangle,
\end{split}
\end{equation}
where we suppress the spatial positions. 
In chaotic theories, at large Lorentzian times $t_1,t_2$, the double commutator is expected to grow exponentially:
\begin{equation}
    \lim_{t_1,t_2\rightarrow \infty} W_R(t_1,t_2) \sim \frac{1}{N} \exp(+\lambda_L (t_1+t_2)/2)\;.
\end{equation}

Since the double commutator can be written in terms of analytically continued $4$-point functions \eqref{eq:conn_four}, \eqref{eq:double_comm} satisfies a ``retarded'' version of the kernel structure of figure  \hyperref[fig:fullkernel]{3a}: 
\begin{equation}\label{eq:double_commutator_kernel}
    W_R=\sum_{n=0}^\infty K_R^n F_0 \ \Longrightarrow\ (1-K_R)W_R = F_0\;,
\end{equation}
with $K_R$ the retarded kernel \cite{KitaevTalk2}. 
The retarded kernel is composed of the same diagrams as in figure \hyperref[fig:fullkernel]{3a}, where one plugs in specific analytic continuations in time of the $n$-point functions.

If $\lambda_L>0$, the ladder structure (however complicated) allows us to compute it in cases where the averaged correlator has conformal symmetry (see \cite{Murugan:2017eto} for a review). This is done by finding the largest solution $\lambda_L^{\text{ker}}$ of the equation $k_R(\lambda_L^{\text{ker}})=1$, where $k_R(\lambda)$ are the eigenvalues of the retarded kernel $K_R$. If $\lambda_L^{\text{ker}}>0$ then we can identify $\lambda_L=\lambda_L^{\text{ker}}$, and otherwise we learn that $\lambda_L=0$.

Importantly, if $J$ is exactly marginal, $k_R(\lambda_L)$ can be found perturbatively in $J$. As a result, one can compute $\lambda_L^{\text{ker}}(J)$ in orders of $J$ by using finitely many core CFT correlation functions at every order. The leading order of the equation is explicitly
\begin{equation}\label{eq:k_in_pert_theory}
\begin{split}
    & k_R(\lambda) = \frac{J^2}{4} \int d^2 x_3 d^2 x_4 \exp(\lambda/2 (t_3+t_4-t_1-t_2)) \\
    &\cdot \langle [\mathcal{O}(\beta/2+i t_2), \mathcal{O}(\beta/2+i t_4)]
    [\mathcal{O}(i t_1), \mathcal{O}(i t_3)]\rangle_0 
    \\
    &\cdot \frac{G_{lr,\Delta(q-1)+\frac{\lambda}{2}}(3,4)}{G_{lr,\Delta+\frac{\lambda}{2}}(1,2)} +O(J^4)\;,
\end{split}
\end{equation}
where $\langle \cdot \rangle_0$ denotes an expectation value at $J=0$ (i.e. of the core CFT). The integration range of the points $3,4$ is over the past light-cone of the points $1,2$ respectively, and $G_{lr,\Delta}$ is the analytically continued cylinder $2$-point function:
\begin{equation}
    G_{lr,\Delta}(1,2) = \frac{1}{\left( 4 
    \cosh\left(\frac{t_{12}-x_{12}}{2}\right) 
    \cosh\left(\frac{t_{12}+x_{12}}{2}\right) 
    \right)^\Delta}.
\end{equation}

The expansion in $J$ can also be used to determine whether the transition into chaos would be continuous or discontinuous (see figure \ref{fig:discontinuouschaos}). This requires finding the sign of $\lambda_L^{\text{ker}}$ in the limit $J= 0^+$. Using the leading contribution to $k_R$ in orders of $J^2$ \eqref{eq:k_in_pert_theory}, it is easy to see that the exponent $\lambda_L^{\text{ker}}(J= 0^+)$ is given by the maximal $\lambda$ for which the integral \eqref{eq:k_in_pert_theory} diverges. If the integral diverges at a positive (negative) value of $\lambda$, we have a continuous (discontinuous) transition into chaos. 

At $J=0$, the kernel vanishes, and so it is not clear that $\lambda_L(J=0)$ is related to chaos. Instead, at $J=0$ we find $N$ decoupled core CFTs, and we expect
\begin{equation}
\begin{split}
    \lim_{t_1,t_2\rightarrow \infty} W_R(t_1,t_2)\mid_{J=0} \;\sim \frac{1}{N} \exp(+\lambda_L^0 (t_1+t_2)/2)
\end{split}
\end{equation}
for some $\lambda_L^0$. Note that the CFTs are decoupled and so $\lambda_L^0$ is a property of a single core CFT. However, we emphasize that it is not a chaos exponent in a single core CFT (as we take $t_1,t_2$ to be larger then any timescale of the core CFT); in fact, for unitary theories in $2d$ it is always non-positive, $\lambda_L^0\le 0$~\cite{Caron-Huot:2017vep}.

Surprisingly, under reasonable physical assumptions it can be shown that $\lambda_L^{\text{ker}}(J=0^+)$ and $\lambda_L^0$ are equal \cite{Berkooz:2021ehv}. This amounts to showing that the integral \eqref{eq:k_in_pert_theory} diverges for $\lambda \le \lambda_L^0$ due to the large (negative) $t_3,t_4$ regime of the integrand. Putting these pieces together, we claim that $\lambda_L^{\text{ker}}(J)$  satisfies
\begin{equation} \label{eq:continuity}
    \lambda_L^{\text{ker}}(J=0^+) = \lambda_L^0.
\end{equation}
The LHS is calculated through the kernel equation $k_R(\lambda)=1$ in the limit $J=0^+$, and the RHS is a property of the core CFT. We can interpret the result by noticing that
$\lambda_L^0$ describes the core CFT double-commutator behavior at large time scales. At arbitrarily weak coupling this behavior seems to control the $1\ll t \ll \log N$ behavior of the disordered theory double-commutator.
The result is striking: in order to determine the type of transition into chaos, it is enough to find (the sign of) $\lambda_L^0$ in a single core CFT. Below we give one example for each type of transition into chaos. In both cases we find that our conjecture \eqref{eq:continuity} holds.

\section{Examples of the Onset of chaos}

\subsection{Disordered generalized free fields}

We now discuss our first class of examples with exactly marginal chaos, which are the disordered generalized free fields in $1$d and in $2$d. Generalized free fields (GFFs) are non-local theories with no energy-momentum tensor, but can be obtained by specific large-flavor limits of local theories, and are good toy-models for more complicated theories. Conformal invariance allows us to set the inverse temperature to be $\beta=2\pi$ in the following.

In $1$d, the generalized free fermion model is called the ``cSYK'' model and was introduced in \cite{Gross:2017vhb}. Explicitly, the action is
\begin{equation}
S=S_0+S_\text{SYK}
\end{equation}
with
\begin{equation}
\begin{split}
S_0&=-\Delta \sum_{i=1}^n  \int d \tau_1 d \tau_2 \chi_i\left(\tau_1\right) \frac{\operatorname{sgn}\left(\tau_1-\tau_2\right)}{\left|\tau_1-\tau_2\right|^{2-2 \Delta}} \chi_i\left(\tau_2\right)\;,\\
S_\text{SYK}&=\frac{i^{\frac{q}{2}}}{q !} \sum_{i_1, \ldots, i_{q}=1}^n  \int d \tau J_{i_1 i_2 \ldots i_{q}} \chi_{i_1} \chi_{i_2} \cdots \chi_{i_{q}}\;.
\end{split}
\end{equation}
Here, $J_{i_1...i_q}$ are Gaussian random variables with variance $\langle J_{i_1...i_q}^2\rangle= \frac{J^2(q-1)!}{N^{q-1}}$. Choosing $\Delta=1/q$, the deformation becomes classically marginal, and it is argued in \cite{Gross:2017vhb} that it is exactly marginal at leading order in $1/N$.

The two- and four-point functions of $\chi_i$ for this model were found in \cite{Gross:2017vhb}. It is simple to extend the results also to the double-commutator. One finds that the corresponding retarded kernel $K_R^{\text{cSYK}}$ has eigenvalues
\begin{equation}\label{eq:GFFs_kR}
k^{\text{cSYK}}_R(\lambda)=\left(\frac{\bar{b}(J)}{\bar b(J\to \infty)}\right)^{q}\frac{\Gamma(3-2 \Delta) \Gamma(2 \Delta+\lambda)}{\Gamma(1+2 \Delta) \Gamma(2-2 \Delta+\lambda)}\;,
\end{equation}
where $\bar b(J)$ solves the equation
\begin{equation}\label{eq:bbar}
\frac{\bar{b}^{q}}{1-2 \bar{b}}=\frac{1}{J^2 \psi(1-\Delta) \psi(\Delta)},\; \psi(\Delta) \equiv 2 \cos (\pi \Delta) \Gamma(1-2 \Delta).
\end{equation}

\begin{figure}
	\centering
	\includegraphics[width=1\linewidth]{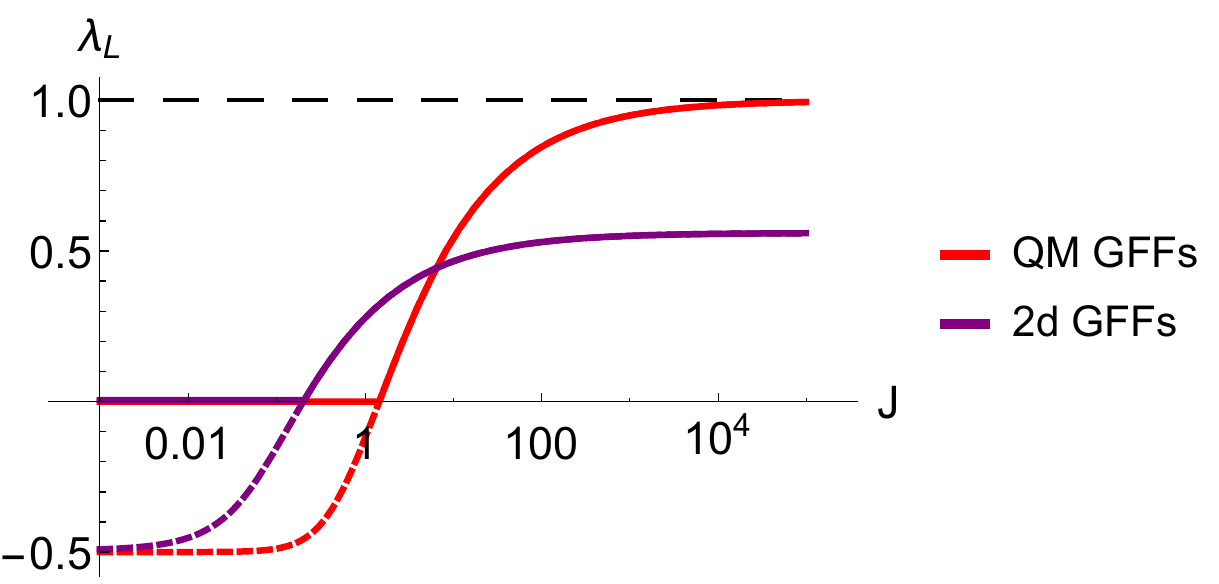}
	\caption{The chaos exponent $\lambda_L(J)$ at $\Delta=1/4$ for disordered GFFs in QM and for SUSY disordered GFFs in $2d$. Dashed lines represent $\lambda_L^{\text{ker}}$ and solid lines represent $\lambda_L$.}
	\label{fig:GFFchaos}
\end{figure}

$\lambda_L^{\text{ker}}$ is found by taking the largest solution to $k^{\text{cSYK}}_R(\lambda)=1$. The result for $\Delta=1/4$ appears in figure \hyperref[fig:GFFchaos]{4}, and similar results apply for other $\Delta$.

At any $\Delta$, $\lambda_L^{\text{ker}}$ approaches the maximal value $\lambda_L=1$ \cite{Maldacena:2015waa} 
at large $J$ as in the SYK model. More relevant to this study, $\lambda_L^{\text{ker}}$ approaches $-2\Delta$ at $J=0$ (corresponding to the dashed red line), and so becomes negative at small $J$ for any $\Delta>0$. As discussed above, we cannot identify $\lambda_L^{\text{ker}}$ with $\lambda_L$ when the former is negative, but we immediately learn that $\lambda_L= 0$ in this regime. We thus conclude that there is no chaos at small enough $J$, corresponding to $\lambda_L$ denoted by the solid red line. We thus find a discontinuous transition into chaos, as in figure \hyperref[fig:discontinuouschaos]{1b}. We also comment that the long time exponent $\lambda_L^0$ of a single core CFT is equal to $\lambda_L^0=-2\Delta$, which matches $\lambda_L^{\text{ker}}(J=0^+)$, and so the conjecture \eqref{eq:continuity} is obeyed. 

The same analysis can be done for disordered GFFs in 2d. In this case we choose to work with $\mathcal{N}=2$ SUSY GFFs, since this results in an exact conformal manifold even at finite $N$. The model consists of $N$ generalized free chiral superfields $\Phi_i$ of dimension $\Delta=1/q$, coupled via the superpotential
\begin{equation}
W=\sum_{i_1\neq ...\neq i_q}^NJ_{i_1...i_q}\Phi_{i_1}...\Phi_{i_q}\;.
\end{equation} 
The computation is very similar, and the results appear in figure \hyperref[fig:GFFchaos]{4}.
At large enough $J$, the chaos exponent for any $\Delta$ approaches the result in the $2$d SUSY versions of the SYK model \cite{Murugan:2017eto,Bulycheva:2018qcp}. In addition, we again find that $\lambda_L^{\text{ker}}$ approaches $-2\Delta$ as $J\to 0^+$, and so for small enough $J$ $\lambda_L^{\text{ker}}$ is negative for any $\Delta>0$. As a result, we again find a discontinuous transition into chaos for any $\Delta>0$. We also find once again that the conjecture \eqref{eq:continuity} is obeyed.

\subsection{Disordered minimal models}

We now discuss the disordered 2d $\mathcal{N}=2$ minimal models. A single core CFT in this case consists of an $\mathcal{N}=2$ SUSY $A_{q-1}$ minimal model, which can be represented by a single chiral superfield $\Phi$ and superpotential $W=\Phi^q$. The full disordered theory is then
\begin{equation}\label{eq:IR_model}
W=\sum_{i=1}^N\Phi_{i}^q+\sum_{i_1\neq....\neq i_q} J_{i_1...i_q}\tilde{\Phi}_{i_1}...\tilde{\Phi}_{i_q}\;.
\end{equation}
We emphasize that we interpret this equation in conformal perturbation theory in $J$ around $N$ copies of the $A_{q-1}$ minimal model, where $\tilde\Phi$ is the chiral operator of dimension $1/q$ which appears in $A_{q-1}$. Since the CFT at $J=0$ has no continuous non-R global symmetries, every classically marginal operator is exactly marginal \cite{Green:2010da,Kol:2002zt,Kol:2010ub}, and so each deformation $J_{i_1...i_q}$ for $i_1\neq...\neq i_q$ is exactly marginal. Thus each realization of the model is conformal, and as a result averaged correlators of $\tilde\Phi_i$ will also be conformal (even at finite $N$).

We can now attempt to compute $\lambda_L(J)$ in perturbation theory. This computation is generically difficult, and so we focus on the case $q=3$, where a core CFT has central charge $c=1$ and so corresponds to the free compact boson at a specific radius. We can identify the operators $\tilde{\Phi}_i$ in terms of vertex operators of the $c=1$ boson, and as a result we can read off all $n$-point functions of the $\tilde{\Phi}_i$. Using this result to compute the leading contribution to the retarded kernel at small $J$ given in \eqref{eq:k_in_pert_theory}, we find that at small $J$ the exponent $\lambda_L^{\text{ker}}$ approaches zero. As a result, we expect a continuous transition into chaos in this model, as in figure \hyperref[fig:discontinuouschaos]{1a}. In particular, the value of $\lambda_L^0$ in a single $A_2$ minimal model is also zero, so that the conjecture \eqref{eq:continuity} is again obeyed.

\begin{acknowledgments}
We would like to thank O. Aharony, N. Brukner, C. Choi, Y. Jia, R.R. Kalloor, B. Lian, J. Maldacena, O. Mamroud, M. Mezei, M. Rangamani, V. Rosenhaus, D. Tong and M. Watanabe for useful conversations.
This work was partly funded by an Israel Science Foundation center for excellence grant (grant number 2289/18), by grant no. 2018068 from the United States-Israel Binational Science Foundation (BSF), by the Minerva foundation with funding from the Federal German Ministry for Education and Research, by the German Research Foundation through a German-Israeli Project Cooperation (DIP) grant ``Holography and the Swampland", and by a research grant from Martin Eisenstein.
MB is the incumbent of the Charles and David Wolfson Professorial Chair of Theoretical Physics.

\end{acknowledgments}

\bibliography{refs}

\end{document}